\documentclass[epsfig,graphics]{article}
\usepackage{amssymb,latexsym,amsmath}
\usepackage{graphics}
\begin{document}
\begin{center}
\huge {\bf Photonuclear interaction of high energy muons and tau-leptons}
\end{center}
\bigskip

\begin{center}E.V. Bugaev, Yu.V. Shlepin \\
\end{center}
\begin{center}Institute for Nuclear Research, the Academy of Sciences  of
 Russia, Moscow, Russia\\
\end{center}
\bigskip

{\bf \noindent Abstract}
{\small General formalism for  the two-component  description of the inelastic lepton-nucleon scattering in the diffractive region is proposed.
Nonperturbative contribution to electromagnetic  structure functions of a nucleon is  described by the modified generalized vector
dominance model containing special cut-off factors restricting the phase volume of the initial $\stackrel{\,~-}{qq}$-pairs of virtual photon's
fluctuations. Perturbative QCD contribution is  described  by  the phenomenological  model suggested (in nonunitarized  form) by
Forshaw, Kerley  and Shaw.

Formulae needed for a numerical  calculation of photonuclear cross sections integrated  over $Q^2$ are presented. It is  argued
that in the case of the photonuclear cross sections at superhigh energies of leptons ($E\gtrsim 10^5GeV$), integrated over 
$Q^2$, the following two-component  scheme is  good  enough: the nonperturbative  contribution  is approximated by the old  parameterization
of Bezrukov and Bugaev, and perturbative one is described by the  model of Forshaw, Kerley  and  Shaw with  parameters determined from  DESY 
data. Corresponding results  of  numerical calculations  of the perturbative part, for the cases of  $\mu A$  and $\tau A$  scattering
at superhigh energies, are given.  }

PACS number(s): 13.60.-r, 12.40.Vv

\section*{\bf \noindent \Large 1.Introduction}

The photonuclear interaction of leptons is , by definition, the process of the inelastic lepton-nucleon or lepton-nucleus scattering,
$$l+N~\to~l+(hadrons).           \eqno(1.1)$$
This process goes through a virtual photon exchange. In the particular case when the four-momentum transferred from the lepton to hadrons is large
($\sqrt{-q^2}>>1GeV$) the process (1.1) is called the deep inelastic scattering; we are interested in the present work in the diffractive
region of the kinematic variables (any $q^2\leqslant 0$, transferred energy $\nu(=-q_0)$ is large, such that $x_{Bj}\sim q^2/2q_0$ is small). This region is especially important
for applications in cosmic ray physics and neutrino astrophysics. The typical problem requiring a knowledge of the photonuclear cross section
is the study of muon and tau propagation through thick layers of matter [1]. High energy muons produced in collisions of cosmic primaries in atmosphere
(the process is $Z_{CR}+Z_{Atm}\to \pi \to \mu$) are detected and studied by modern large underground (and underwater) telescopes.
As for taus, they can be produced in the Earth by extragalactic high energy $\tau-$neutrinos (such neutrinos must always present in an extragalactic
neutrino flux, due to $\nu_{\mu}\longleftrightarrow\nu_{\tau}$ oscillations).

The process (1.1) is reduced to the absorption of the virtual photon by the nucleon,
$$\gamma^{\ast}+N\to(hadrons),      \eqno(1.2)$$
and, by optical theorem, is connected with the Compton scattering of a virtual photon,
$$\gamma^{\ast}+N\to\gamma^{\ast}+N.    \eqno(1.3)$$
The Compton scattering in the diffractive region is described by the vacuum exchange which, in turn, is modelled in QCD by the exchange
of two or more gluons in a colour singlet state.
It becomes possible because, in laboratory system, the interaction region has a large longitudinal size, and the photon developes an internal
structure due to its coupling to quark fields. In the diffractive region, i.e. at small $x_{Bj}$, the diffractive $\gamma^{\ast}N-$scattering is 
dominant in the Compton amplitude, as compared to the direct contribution due to a photon's bare component (although, the direct
process cannot be neglected absolutely, especially at high energies of leptons).

These general statements (about diffractive, hadron-like, nature of $\gamma^{\ast}N-$ interaction and even about the two-gluon exchange) are not
enough for a quantitative calculation of the Compton amplitude or $\gamma^{\ast}N$ total cross section. One must know also the photon
wave function or, by other words, the spectrum of hadron-like constituents of the photon and, besides, the amplitudes of the scattering of these 
constituents on a nucleon. In pre-QCD era these problems were considered, almost exclusively, in terms of vector dominance models.
These models use a hadronic basis for a description of the spectrum of photon's constituents. It is well known, however, that they can describe the main
features of the $\gamma^{\ast}N-$interaction (e.g., the Bjorken scaling) only by using rather unnatural assumptions on hadronic amplitudes
(see, e.g., [2]).
With advent of QCD it became clear that the description of the photon wave function in terms of purely hadronic constituents is not consistent:
the $\stackrel{\,~-}{qq}$-component interacting with a target perturbatively must be added. Such a picture arises due to inherent QCD properties 
(asymptotic freedom, colour transparency).

The aim of the present work is twofold. In the first part of the work we formulate the two-component model of the electromagnetic structure functions of a
nucleon separating, in the general form, the nonperturbative part which is described by a generalized vector dominance model. The difference
with previous works on the subject [3,4] consists just in the use in our approach the generalized vector dominance, without limiting oneself by low-mass
vector mesons ($\rho,\omega,\phi$) only. At the second part of the work we consider the perturbative component and carry out the concrete
calculations of photonuclear cross sections for leptons of superhigh energies.

The paper is organized as follows. In sec.2 the general formulas for the two-component structure functions are derived. The hard (nonperturbative)
component of the structure functions is considered, in a framework of the colour dipole model, in Sec.3. In Sec.4. we show how the 
existing widely used formula [5]  for the photonuclear cross section must be modified to take into account the contribution of the 
hard   component. Results of numerical calculations and conclusions are presented in the last section.

\section*{\bf \noindent \Large 2.Two-component model of structure functions}

\subsection*{\bf A.Two-step picture of the Compton scattering}

The starting  point of  our  consideration    is the  expression   based  on   the perturbative QCD  and the  two-step picture
of the $\gamma^\ast  p$ interaction: the first step 
is the $\gamma^\ast  \rightarrow \stackrel{\,~-}{qq}$ conversion, and the second step is an interaction of the $\stackrel{\,~-}{qq}$-pair
with the target proton. The total $\gamma^\ast p$ interaction cross section ( summed over all possible final hadronic states) 
is given by the contribution of the $\stackrel{\,~-}{qq}$-channel in the imaginary part of the Compton forward scattering amplitude 
( fig.1):
$$\sigma_{T,L}(Q^2,s)=\frac{1}{16\pi^3}\sum\limits_{f}\sum\limits_{r,r'}\int dz\int d^2k_\bot \int dz' \int d^2k'_\bot\times$$
$$\times {\psi^{T,L}_{\gamma(r,r')}}^\ast(\stackrel{\to}{k}'_\bot,z',Q^2)\frac{1}{s}\mathcal{A}_{\stackrel{\,~-}{qq}+p}
(\stackrel{\to}{k}'_\bot ,z',\stackrel{\to}{k}_\bot,z;s) \psi^{T,L}_{\gamma(r,r')}(\stackrel{\to}{k}_\bot,z,Q^2). 
                       \eqno(2.1) $$
In this expression $\mathcal{A}_{\stackrel{\,~-}{qq}+p}$ is the imaginary part of the $(\stackrel{\,~-}{qq})p$ forward scattering amplitude,
$Q^2$ and $\sqrt{s}$ are a virtuality of the photon and a $\gamma^\ast p$ center-of-mass energy, $\psi_{\gamma(r,r')}$ is the light cone 
wave function of  $\stackrel{\,~-}{qq}$-fluctuations of a virtual photon with transverse or longitudinal polarization. 
This wave function depends on quark and antiquark helicities $(r,r')$, quark mass $(m_f)$ and quark momentum variables
$z,k_\bot$ ($z=k_+/q_+$ is the fraction of the photon light cone momentum carried by the quark, $k_\bot$ is the transverse
momentum of the incoming quark). In the following we will need the expressions for the squares of these wave functions summed
over the quark helicities:
$$\sum\limits_{r,r'}{\psi^{T}_{\gamma(r,r')}}^\ast(\stackrel{\to}{k}'_\bot,z,Q^2)\psi^{T}_{\gamma(r,r')}(\stackrel{\to}{k_\bot},z,Q^2)
=2N_c\frac{e_f^2(\stackrel{\to}{k'_\bot}
\stackrel{\to}{k_{\bot}}[z^2+{(1-z)}^2]+m_f^2)}{(M^2+Q^2) ({M'}^2+Q^2) z^2 {(1-z)}^2}, $$
$$\sum\limits_{r,r'}{\psi^{L}_{\gamma(r,r')}}^\ast(\stackrel{\to}{k}'_\bot,z,Q^2)\psi^{L}_{\gamma(r,r')}(\stackrel{\to}{k_{\bot}},z,Q^2)=
8N_c\frac{e_f^2Q^2}{(M^2+Q^2)({M'}^2+Q^2)}.          \eqno(2.2)$$
In these formulas $M$ and $M'$ are invariant masses of the incoming and outcoming  $\stackrel{\,~-}{qq}$-pairs, 
$$M^2=\frac{k_\bot^2+m^2_f}{z(1-z)},~~{M'}^2=\frac{{k'}^2_\bot+m^2_f}{z(1-z)}.
                            \eqno(2.3)$$

\subsection*{\bf B.Two gluon exchange approximation}

We suppose,   further, that the forward scattering amplitude in Eq.(2.1) can be written in the form
$$\frac{1}{s}\mathcal{A}_{\stackrel{\,~-}{qq}+p}(\stackrel{\to}{k}'_\bot,z',\stackrel{\to}{k}_\bot,z;s)=\{\delta(\stackrel{\to}{k}'_\bot-
\stackrel{\to}{k_{\bot}})\int d^2l'_\bot B({l'}^2_\bot,s)-B(l^2_\bot,s)\}\delta(z'-z),                    \eqno(2.4)$$
$$\stackrel{\to}{l}_\bot=\stackrel{\to}{k}'_\bot-\stackrel{\to}{k_{\bot}}.   \eqno(2.4a)$$

Such a form of the $(\stackrel{\,~-}{qq})p$ amplitude is suggested by the model of the structure functions [6] based on the two-gluon exchange 
approximation of perturbative  QCD [7-10].
The first   term in Eq.(2.4)   corresponds to the   diagonal transition when $M^2={M'}^2$; the delta-function factor in  this term 
arises due  to  zero transverse momentum  transfer to the  quark (  at high  energies the value of   $z$ is frozen during the  scattering 
 process, and the interaction can, in a general case,  change  only the transverse  momentum of $\stackrel{\,~-}{qq}$ pair's
 particles ). In the  case  of nondiagonal transitions ( the second term in Eq.(2.4)) the  scattering amplitude depends
 on   the square  of the  transverse momentum,  $
\stackrel{\to}{l}_\bot=\stackrel{\to}{k}'_\bot-\stackrel{\to}{k_{\bot}}$,
 transferred to the quark.
 
 In spite of  the  fact that the  expressions (2.1), (2.4) are based on the perturbation theory, they can  be  used for an  
 approximated calculation  of the photoabsorption cross section and  electromagnetic structure   functions of a nucleon at low
 $Q^2$, where  nonperturbative  effects are important. It is assumed in such   an  approach  that  the  perturbative 
 fluctuation of the  $\gamma^\ast$ into  a $\stackrel{\,~-}{qq}$ pair is the  dominant  process and the subsequent colour 
 singlet exchange  is  realized by  two gluons only.  The problem of  the  inclusion   of higher Fock states 
 (i.e., $\stackrel{\,~-}{qq}g$ ) in  the photon   wave function   was discussed  in   the  third   work   of ref.[6]. As for the
 two-gluon exchange appoximation for  a description of   the  elastic   $(\stackrel{\,~-}{qq})p$ - amplitude, the  nonperturbative  effects
 are  approximately  taken into account  by introducing an  effective mass of the gluons and an effective quark-gluon  coupling
 constant.   The description of soft hadronic processes via the exchange of two effective gluons was suggested many years ago
 in  the  works  [7,8].
 
 Below we will use the simple ansatz  for the function $B(l^2_\bot,s)$ at small $l^2_\bot$:
$$B(l^2_\bot,s)=\frac{\sigma_0(s)}{l^2_\bot}.                           \eqno(2.5)$$

The $l_{\bot}$ dependence of this type is obtained in the models of refs.[6-10] 
 if, in addition to the two-gluon exchange, a constituent quark
model of the target nucleon is used. More exactly, in this case one has [6] 
$$B(l^2_\bot,s)=\frac {V(\stackrel{\to}{l_{\bot}})}{{(l^2_\bot+\mu_G)}^2},  \eqno(2.6)$$
\noindent 
where $V(\stackrel{\to}{l_{\bot}})$ is the GGNN-vertex function,    
$$V(\stackrel{\to}{l_{\bot}})\cong 1-e^{-\frac{l^2_\bot}{2<r^2_N>}},          \eqno(2.7)$$

\noindent $\mu_G$ is the effective mass of the gluon, $<r^2_N>$ is the mean square radius of a nucleon. If  $\mu_G$ is small,
 there is the region,
$$\mu^2_G<<l^2_\bot\stackrel{<}{~}\frac{1}{<r^2_N>},               \eqno(2.8)$$
in which $B(l^2_\bot,s)\sim 1/l^2_\bot$, as in Eq.(2.5).

In purely perturbative QCD (colour dipole picture of the interaction [6,10-12], leading $\alpha_s ln\frac1x$ approximation)
 one obtains for the $B(l^2_\bot,s)$ function:
$$B(l^2_\bot,s)\cong \frac{4\pi}3 f(x',l^2_\bot)\alpha_s(l^2_\bot)\frac1{l^4_\bot},  \eqno(2.9)$$ 
$$x'=\frac{Q^2+M^2}{s}.      \eqno(2.9a)$$
Here, $f(x,l^2_\bot)$ is the unintegrated gluon distribution in a proton. One can conclude, comparing Eqs.(2.5) and (2.9), that
the ansatz(2.5) is not justified by the perturbative QCD calculation.

\subsection*{\bf C.Separation of perturbative and nonperturbative parts}

Below, in this section, we use the general expressions (2.1-2.4) for a  separation of soft (nonperturbative) and hard (perturbative)
contributions in the structure functions.

A nature of the $(\stackrel{\,~-}{qq})p$ interaction is connected with a transverse size of the $\stackrel{\,~-}{qq}$-pair:
only for a small $r_\bot$ this nature is, due to the asymptotic freedom, perturbative while the pairs with a large enough $r_\bot$
interact nonperturbatively. An average transverse separation between particles of the pair is
$$\stackrel{-}{r}_\bot=v_{\bot,relative}\tau,               \eqno(2.10)$$
where $v_{\bot,relative}$ is an relative velocity in the transverse direction,

$$v_{\bot,relative}=\frac12(\frac{k_\bot}E+\frac{k_\bot}{E'}),$$
$$E=\sqrt{zk_\gamma+k^2_\bot+m^2_f},~ E'=
\sqrt{(1-z)k_\gamma+k^2_\bot+m^2_f},  \eqno(2.11)$$
and $\tau$ is the lifetime of the $\stackrel{\,~-}{qq}$-fluctuation,
$$\tau=\frac1{\sqrt{k^2_\gamma+M^2}-\sqrt{k^2_\gamma-Q^2}}\cong\frac{2k_\gamma}{M^2+Q^2},       \eqno(2.12)$$
where $k_\gamma$ is the virtual photon's three-momentum.

If follows from Eqs.(2.10-2.12) that
$$\stackrel{-}{r}_\bot\cong\frac1{k_\bot(1+\frac{Q^2}{M^2})}.      \eqno(2.13)$$

We use the following criterion of the nonperturbative nature of the interaction: the $(\stackrel{\,~-}{qq})p$ interaction is nonperturbative if 
$$\stackrel{-}{r}_\bot\geqslant r_{\bot 0}\equiv  \frac 1{k_{\bot 0}},    \eqno(2.14)$$
where $k_{\bot 0}$ is some parameter. This criterion, together with Eqs.(2.3) (where the approximation $m_f=0$ is used)
and Eq.(2.13), leads to the corresponding constraint for the variable $z$,
$$z(1-z)\leqslant \frac{k^2_{\bot 0}M^2}{(M^2+Q^2)^2}= 
\begin{cases}
\frac {k^2_{\bot0}}{M^2},~ Q^2<<M^2  \\
~~~~~~~~~~~~~~~~~~~~~\\
\frac{k^2_{\bot 0}M^2}{Q^4},~Q^2>>M^2,
\end{cases}          \eqno(2.14a)$$
according to which the $(\stackrel{\,~-}{qq})p$ interaction is nonperturbative if the $\stackrel{\,~-}{qq}$-pair is
asymmetric [13], i.e., $E \ll E'$ or $E'\ll E$. In turn, the high degree of asymmetry corresponds (in accordance with Eqs.(2.3)),
at a given value of $M$, to a small $k_\bot$ of pair's particles and to their alignment in the 
direction of the virtual photon (the "aligned jet" conjecture had been suggested in the work [14]).

\subsection*{\bf D.Hard component}

It is convenient, for considering the structure functions in the perturbative domain, to use the impact parameter representation.
The resulting formula for $\sigma_{T,L}(Q^2,s)$, which follows from Eqs.(2.1-2.4), is well known [6]:
$$\sigma_{T,L}(Q^2,s)=\int dz\int d^2r_\bot\sigma(r_\bot,s){\mid \psi^{T,L}_\gamma(\stackrel{\to}{r}_\bot,z,Q^2)\mid}^2.    \eqno(2.15)$$

The squares of the photon wave functions in impact parameter space summed over helicities are given by the  expressions
$$\mid\psi^T_\gamma(r_\bot,z,Q^2)\mid^2=\frac{3\alpha}{2\pi^2}\sum\limits_f{(\frac{e_f}e)}^2\{[z^2+{(1-z)}^2]\stackrel{-~}{Q^2}K^2_1
(\stackrel{-}{Q}r_\bot)+m^2_fK^2_0(\stackrel{-}{Q}r_\bot)\},$$
$$\mid\psi^L_\gamma(r_\bot,z,Q^2)\mid^2=\frac{3\alpha}{2\pi^2}\sum\limits_f{(\frac{e_f}e)}^24z^2{(1-z)}^2Q^2K^2_0(\stackrel{-}{Q}r_\bot)\},  
      \eqno(2.16)$$
$$\stackrel{-~}{Q^2}\equiv Q^2z(1-z)+m^2_f.                       \eqno(2.16a) $$

The factor $\sigma(r_\bot,s)$ in Eq.(2.15) is a total cross section of the interaction of the $\stackrel{\,~-}{qq}$-pair of the 
transverse size $r_\bot$ with a target proton,
$$\sigma(r_\bot,s)=\int d^2l_\bot(1-e^{-i\stackrel{\to}{l}_\bot\stackrel{\to}{r}_\bot})B(l^2_\bot,s).                        \eqno(2.17)$$
Using Eq.(2.9) for $B(l^2_\bot,s)$, one obtains (for small values of $r_\bot$) 
$$\sigma(r_\bot,s) \approx\frac{\pi^2}3r^2_\bot\alpha_s(\frac{1}{r^2_\bot})x'_{eff}G(x'_{eff},\frac1{r^2_\bot}),   \eqno(2.18)$$ 
where $xG(x,Q^2)$ is the gluon structure function of a proton.
Approximately one has 
$$x'_{eff}\approx\frac{{(k^2_\bot)}_{eff}}{s}\sim\frac1{r^2_\bot s},               \eqno(2.19)$$
and[12]
$$\sigma(r_\bot,s)\approx\frac{\pi^2}{3}r^2_\bot\alpha_s(\frac1{r^2_\bot}){[xG(x,\frac1{r^2_\bot})]}_{x=\frac1{r^2_\bot s}}.    \eqno(2.20)$$

Below we will use, instead of Eq.(2.20), the phenomenological model [15] for $\sigma(r_\bot,s)$, containing except the colour transparency 
factor $r^2_\bot$ also the cut-off factor at large $r_\bot$ (which is necessary for a separation of the perturbative part of the structure 
functions). In such models one has
$$\sigma(r_\bot,s)\sim r^2_\bot e^{-\frac{r_\bot}{r^0_\bot}}(r^2_\bot s)^{\lambda_H}.    \eqno(2.21)$$

Note, that $s$-variable enters the r.h.s. of Eq.(2.21) in combination with $r_\bot^2$ only, i.e., in the same manner as in
the formula (2.20) obtained in perturbative QCD.

\subsection*{\bf E.Soft  component}

For considering the nonperturbative part of $\sigma_{T,L}(Q^2,s)$ we rewrite Eq.(2.1) in a form of the double 
dispersion relation. For this aim we perform the following change of variables:
$$\int dz\int d^2k_\bot\int d^2k'_\bot=$$
$$=\pi\int dzz^2(1-z)^2\int dM^2\int d{M'}^2\int\frac{{dl^\ast_\bot}^2}{2MM'\sqrt{1-cos^2\phi}}, \eqno(2.22)$$
$$cos^2\phi=\frac1{4M^2{M'}^2}(M^2+{M'}^2-{l^\ast_\bot}^2),       \eqno(2.22a)$$
$${l^\ast_\bot}^2\equiv\frac{l^2_\bot}{z(1-z)},~~~~~~~~(M-M')^2\leq{l^\ast_\bot}^2\leq(M+M')^2.        \eqno(2.22b) $$
\noindent 
Using this change of variables and the expressions (2.2) for the squares of the photon wave functions we have, instead of Eq.(2.1),
$$\sigma_{T,L}(Q^2,s)=\sum\limits_f\int\frac{dM^2}{M^2+Q^2}\times$$
$$\times\int\frac{d{M'}^2}{{M'}^2+Q^2}\int dz \rho_{T,L}(z)\frac1{s}
ImA^{T,L}_{\stackrel{\,~-}{qq}+p}(M^2,{M'}^2,s),  \eqno(2.23)$$
$$\rho_T(z)=\frac32[z^2+(1-z)^2],    \eqno(2.23a)$$
$$\rho_L(z)=\frac324z(1-z).    \eqno(2.23b)$$
The functions $\rho_{T,L}(z)$ are normalized densities of $\stackrel{\,~-}{qq}$ states in the $z$-space. 
Imaginary parts of the forward  $(\stackrel{\,~-}{qq})p$ scattering amplitudes $A^{T,L}_{\stackrel{\,~-}{qq}+p}(M^2,{M'}^2,s)$ are  obtained 
using  Eqs.(2.4,2.5):
$$\frac1s ImA^{T}_{\stackrel{\,~-}{qq}+p}(M^2,{M'}^2,s)=\frac1{12\pi^2}N_ce^2_f\{M^2\delta(M^2-{M'}^2)[\int d^2l'_\bot B({l'}^2_\bot,s)]-$$
$$-\int\frac{d{l^\ast_\bot}^2}
{2\sqrt{1-cos^2\phi}}B({l^\ast_\bot}^2,s)\},                            \eqno(2.24)$$
$$\frac1s ImA^{L}_{\stackrel{\,~-}{qq}+p}(M^2,{M'}^2,s)=\frac{Q^2}{MM'}\frac 1sImA^{T}_{\stackrel{\,~-}{qq}+p}(M^2,{M'}^2,s).  \eqno(2.24a)$$
For  convenience we  included  the  factor $\frac{Q^2}{MM'}$ in  the  definition of a  longitudinal $(\stackrel{\,~-}{qq})p$ amplitude $A_L$.

It is important to underline that for the derivation of Eqs.(2.23-2.24) the expression (2.5) for the function $B(l^2_\bot,s)$ was used.
The factorization of a $z$-dependence and the simple form of this dependence given by the functions of Eqs.(2.23a,b) became possible
just due to the assumption that $B(l_{\bot}^2,s)\sim l_{\bot}^{-2}$. The use of this assumption is leqitimate if we consider
the nonperturbative component of the structure functions.

Now, for the separation of the nonperturbative contribution in the double dispersion relation (2.23)
one must take into account the restriction  of  the  $\stackrel{\,~-}{qq}$ pair's  phase space  due to the constraint  (2.14a).
After such a  restriction the  nonperturbative part of $\sigma_{T,L}(Q^2,s)$  can be written in   the  form
$$\sigma_{T,L}(Q^2,s)=\sum\limits_f\int\frac{dM^2}{M^2+Q^2}\times$$
$$\times\int\frac{d{M'}^2}{{M'}^2+Q^2}\eta_{T,L}(M^2,{M'}^2,k_{\bot 0}^2)\frac1{s}
ImA^{T,L}_{\stackrel{\,~-}{qq}+p}(M^2,{M'}^2,s).  \eqno(2.25)$$

The cut-off functions  $\eta_{T,L}(M^2,{M'}^2,k_{\bot 0}^2)$ are   obtained in the  Appendix A. In the   region 
$M^2,{M'}^2>>k_{\bot 0}^2,Q^2$  they are  given  by the  simple  expressions:  
$$\eta_T(M^2,{M'}^2,k_{\bot 0}^2)\sim 3\frac{k_{\bot 0}^2}{MM'};~~~
~~\eta_L(M^2,{M'}^2,k_{\bot 0}^2)\sim 6\frac{k_{\bot 0}^4}{M^2{M'}^2}.        \eqno(2.26)$$

The  dispersion  relation (2.25) can be rewritten by introducing (somewhat artificially) the  discrete basis for a
description of invariant mass spectra of $\stackrel{\,~-}{qq}$ pairs: 
$$\sigma_{T,L}(Q^2,s)=\sum\limits_f\int\frac{M^2dM^2}{M^2+Q^2}\times$$
$$\times\int\frac{{M'}^2d{M'}^2}{{M'}^2+Q^2}\rho_{T,L}(M^2,{M'}^2,k_{\bot 0}^2)\frac1{s}
ImA^{T,L}_{\stackrel{\,~-}{qq}+p}(M^2,{M'}^2,s),  \eqno(2.27)$$
$$\rho_{T,L}(M^2,{M'}^2,k_{\bot 0}^2)=
\sum\limits_{n,n'}\eta_{T,L}(M^2,{M'}^2,k_{\bot 0}^2)\delta(M^2-M_n^2)\delta({M'}^2-M_{n'}^2).         \eqno(2.28)$$
Here, $\rho_{T,L}(M^2,{M'}^2,k_{\bot 0}^2)$   are spectral functions  giving densities of  $\stackrel{\,~-}{qq}$  -  states
of  definite  masses.  

\subsection*{\bf F.Generalized vector dominance approach for the soft component}

For a quantitative calculation of the nonperturbative contribution to the structure functions we use the approach based on generalized vector dominance
(GVD) ideas [16,17]. According to hte GVD approach, a $\stackrel{\,~-}{qq}$-pair's quantum state which is modified by strong interactions of quark
and antiquark can be presented by the sum of hadronic (vector meson) states with fixed masses,
$$\mid\stackrel{\,~-}{qq}^{free},M>~~\rightarrow~~\mid\stackrel{\,~-}{qq}^{int.},M>=\sum\limits_n\frac{e}{f_n}M^2_n
\delta(M^2-M^2_n)\mid V_n,M_n>.     \eqno(2.29)$$
Further, it is assumed that the vector mesons are those observed in $e^+e^-$ annihilation process and, correspondingly, the 
photon-meson coupling constants $\frac e{f_n}$ (i.e., the coefficients in the sum in Eq.(2.29)) are expressed through the 
vector meson's leptonic widths.

Following the experience of  vector dominance  models [18,19],  we  will distinguish  vector   mesons of different  polarizations.
We assume  that  the  meson's  polarization  correlates  with a polarization  of  the initial virtual photon.

We  can   write now the basic  double dispersion  relation of GVDM  [16,17], in  which  the  cut variables,  $M$ and  $M'$,
are  the  masses of the  incoming and outgoing  vector  meson states  in   the quasi-elastic forward  amplitudes  $T^{T,L}$  of
the  vector  meson-proton  scattering:  
$$\sigma_{T,L}(Q^2,s)=\int\frac{M^2dM^2}{M^2+Q^2}\int\frac{{M'}^2d{M'}^2}{{M'}^2+Q^2}\times$$
$$\times\rho^{GVD}_{T,L}(M^2,{M'}^2,s)\frac{e}{f(M^2)}\frac{e}{f({M'}^2)}\frac1{s}ImT^{T,L}(M^2,{M'}^2,s),  \eqno(2.30)$$
and the spectral function $\rho^{GVD}_{T,L}$  is a density of the vector meson states.
 The connection between 
transverse  and  longitudinal  amplitudes  is  
$$\frac1s ImT^{L}(M^2,{M'}^2,s)=\xi\frac{Q^2}{MM'}\frac 1sImT^{T}(M^2,{M'}^2,s),  \eqno(2.31)$$
where parameter $\xi$ takes into account the possible difference in interactions of vector mesons with different polarizations.
The  normalization of these amplitudes  is   given    by   the relations
$$\frac 1s ImT^{T}(M_n^2,M_n^2,s)=\sigma_{V_np}(s),                          \eqno(2.32a)$$
$$\frac 1s ImT^{L}(M_n^2,M_n^2,s)=\frac{Q^2}{M_n^2}\xi \sigma_{V_np}(s).      \eqno(2.32b)$$

One  must note that the introducing of the factor  $\xi$ in  Eq.(2.31) is  at  variance (if  $\xi<1$)  with the initial two-step 
formula  (2.1)  based  on   the assumption  that  there is  the strict  factorization (i.e.,  independence  of  a 
$(\stackrel{\,~-}{qq})p$ amplitude on  the virtual photon's polarization ).  

The crucial assumption of   our  GVD approach consists  in  the following: the $M^2$- and ${M'}^2$- dependencies of the spectral functions  
$\rho^{GVD}_{T,L}$ and $\rho_{T,L}$ (Eq.(2.28))  are the same, i.e., 
$$\rho^{GVD}_{T,L}(M^2,{M'}^2,s)=\rho_{T,L}(M^2,{M'}^2,k_{\bot  0}^2).  \eqno(2.33)$$

The physical sense of this  equality   is  evident: the hadronic   basis is applicable for a description of the spectrum  
of virtual  photon fluctuations if (and only if) $\stackrel{\,~-}{qq}$ pairs produced in the $\gamma^*\to\stackrel{\,~-}{qq}$
transition interact with the  target proton   nonperturbatively. The separation of the nonperturbative component leads to
the restriction of the phase space of $\stackrel{\,~-}{qq}$ pairs and, in turn, to the corresponding constraint on the 
spectral function, i.e., on the density of the vector meson states. 

Inserting Eq.(2.33) into Eq.(2.30) one finally obtains 
$$\sigma_{T,L}(Q^2,s)=\sum\frac{e^2}{f_nf_n'}\frac{M_n^2}{M_n^2+Q^2}\frac{M^2_{n'}}{M^2_{n'}+Q^2}\eta_{T,L}(M_n^2,M^2_{n'},k_{\bot 0}^2)
\frac1{s}ImT^{T,L}_{n,n'}(s).  \eqno(2.34)$$
This expression differs from a standard GVDM formula only by the presence of the cut-off factors $\eta_{T,L}(M_n^2,{M}^2_{n'},k_{\bot 0}^2)$, 
the expressions for which are given above. These cut-off factors, and the nonperturbative contribution to the structure functions given
by Eq.(2.34), depend on the parameter $r_{\bot 0}(\equiv 1/{k_{\bot 0}})$. 

In contrast with the two-component models developed in the works [3,4] our treatment of the nonperturbative part contains no restrictions on 
the vector meson
 mass spectrum: the masses of vector mesons may be arbitrarily large. At the same time, due to a presence of the cut-off
 factors, it needs no special cancellation mechanism (consisting, e.g.,in an   introducing of unphysically large nondiagonal terms )
for ensuring a convergence of the sums in Eq.(2.34) [2]. 

The  further consistent  developing of the two-component   model outlined here (first of  all, the   concrete realization of the modified  
GVDM based  on  Eq.(2.34)) is a subject  of the next  work  of the  present   authors.   Below, in Sec.5,  we  suggest the  simplified  
approach  for  practical  calculations of photonuclear cross sections at very high lepton energies ($\gtrsim 10^5GeV$),
integrated over $Q^2$,  in which the nonperturbative  component  of the structure  functions is appoximated by the known formulae based
on the standard  (nondiagonal) GVDM working rather well   at lepton energies $\lesssim 10^5GeV$.   

\section*{\bf \noindent \Large 3.Hard component}

For numerical caculations of the hard component we use the analytical form for the function $\sigma(r_\bot,s)$ suggested in ref.[15]. Namely,
$$\sigma(r_\bot,s)=(a_1r_\bot+a_2r_\bot^2+a_3r_\bot^3)^2e^{-\frac{r_\bot}{{r_\bot^0}}}(r_\bot^2 s)^{\lambda_H}.     \eqno(3.1)$$
The numerical values of the coefficients $a_1,a_2,a_3,r_\bot^0,\lambda_H$ were found by authors of work [15] by fitting the ZEUS data.
The main practical aim of our work is the calculation of the cross section of the inelastic lepton 
(muon and tau) scattering at extremely high energies (up to $10^9GeV$). Correspondingly, we need structure functions $\sigma(Q^2,s)$ 
at very low values of Bjorken variable $x\cong Q^2/s$ and, therefore, need extrapolation of HERA data to these values.
The Regge-type $s$-dependence of  $\sigma(r_\bot,s)$ in Eq.(3.1) is definitely incorrect at very high energies due to the 
unitarity constraint.
So, we are forced to use additional assumptions connected with an unitarization procedure.
The simplest way of the unitarization is the following (see, e.g., [20]). An amplitude for the elastic scattering of a $\stackrel{\,~-}{qq}$-pair
in impact parameter space which is connected with the cross section $\sigma(r_\bot,s)$ by the relation
$$\sigma(r_\bot,s)=2\int d^2b_\bot Ima(r_\bot,b_\bot,s),               \eqno(3.2)$$
is, by assumption, purely imaginary at high energies and can be expressed through the opacity function $\Omega$:
$$a(r_\bot,b_\bot,s) =i\{1-e^{\frac 12\Omega(r_\bot,b_\bot,s)}\}.        \eqno(3.3)$$
The $b_\bot$-dependence of the opacity function $\Omega$ cannot be determined 
without a concrete model. We assume the form [20]
$$\Omega(r_\bot,b_\bot,s)=\sigma(r_\bot,s)S(b_\bot^2),          \eqno(3.4)$$
where the profile function $S(b_\bot^2)$ is given by the formula [20]
$$S(b_\bot^2)=\frac1{\pi R^2}e^{-\frac{b_\bot^2}{R^2}},       \eqno(3.5)$$
$$R^2\cong 10GeV^2.         \eqno(3.5a) $$
The unitarized cross section is given by the integral
$$\sigma^{unit}(r_\bot,s)=2\int
d^2b_\bot(1-e^{-\frac12\Omega(r_\bot,b_\bot,s)})=$$
$$=2\pi R^2\left\{ln\frac{\sigma(r_\bot,s)}{2\pi R^2}+C-
Ei(-\frac{\sigma(r_\bot,s)}{2\pi R^2})\right\},  \eqno(3.6)$$
where $C=0.577$.

Asymptotically
$$Ei(-x)\mid_{x\to\infty}\sim\frac1xe^{-x},      \eqno(3.7)$$
so, in the limit of high energies
$$\sigma^{unit}(r_\bot,s)\sim2\pi R^2\lambda_H lns,              \eqno(3.8)$$
if the $s$-dependence of $\sigma(r_\bot,s)$ is $\sim s^{\lambda_H}.$

\section*{\bf \Large \noindent 4.Cross section of photonuclear interaction}

The differential cross section of the inelastic lepton-nucleon scattering is expressed through the electromagnetic structure functions
$W_1,W_2$ of nucleons:
$$\frac{d^2\sigma}{d\nu dQ^2}=\frac{2\pi\alpha^2}{Q^4}\frac1{E^2}\{(Q^2-2m^2_l)W_1(Q^2,\nu)+$$
$$+(2E(E-\nu)-\frac{Q^2}{2})W_2(Q^2,\nu)\},                      \eqno(4.1)$$
where $E$ is an lepton energy in laboratory system and $m_l$ is the lepton mass. The variable $\nu$ in Eq.(4.1) is related with 
the $\gamma^{\ast}p$ center-of-mass energy $\sqrt{s}$ used above,
$$\nu=\frac{Q^2}{2M_p}x_{Bj}^{-1}=\frac{s+Q^2-M^2_p}{2M_p}\approx\frac{s}{2M_p}.    \eqno(4.2)$$
The functions $W_1,W_2$ are connected with the structure functions $\sigma_{T,L}$ by the simple relations:
$$\sigma_T(Q^2,\nu)=\frac{4\pi^2\alpha}{K}W_1(Q^2,\nu),$$
$$\sigma_T(Q^2,\nu)+\sigma_L(Q^2,\nu)=\frac{4\pi^2\alpha}{K}\frac{Q^2+\nu^2}{Q^2}W_2(Q^2,\nu),  \eqno(4.3)$$
$$K=\nu-\frac{Q^2}{2M_p}\approx\nu.       \eqno(4.3a) $$

For problems of cosmic ray physics and high energy particle physics one 
needs the lepton-nucleon inelastic scattering cross
section integrated 
over $Q^2$. Introducing the new variable $v=\frac {\nu}{E}$ 
(which is the fraction of the lepton energy 
transferred  to newly produced particles) we obtain

$$v\frac{d\sigma}{dv}(E,v,m_l)=\frac{\alpha}{2\pi}\int\limits^{Q_{max}^2}_{Q^2_{min}}\frac{dQ^2}{Q^2}\times
\frac{1-\frac{Q^2}{2M_pvE}}{1+\frac{Q^2}{(vE)^2}}\times$$
$$\times\left\{ v^2\sigma_T[1+\frac{Q^2}{(vE)^2}](1-\frac{2m^2_l}{Q^2})+2(1-v)(\sigma_T+\sigma_L)\right\},  \eqno(4.4)$$
$$ Q^2_{min}\simeq\frac{m^2_lv^2}{1-v},~~~~~~Q_{max}^2=2M_pvE.           \eqno(4.4a)$$
As it follows from Eq.(4.3), in the region of small $x_{Bj}$
$$\sigma_L(Q^2,\nu)+\sigma_T(Q^2,\nu)\sim\nu W_2(Q^2,\nu)\frac{1}{Q^2}\sim\frac{1}{Q^2}, \eqno(4.5)$$
therefore the contribution of small $Q^2$ (near $Q^2_{min}$) dominates in the integral in Eq.(4.4). So, 
as far as we are interested in $Q^2-$integrated values (it is just the case of cosmic ray physics) we can
safely use for a theoretical prediction of the structure functions $\sigma_{T,L}(Q^2,\nu)$ the models working
well just in the small $Q^2-$region. In particular, the structure functions $\sigma_{T,L}(Q^2,\nu)$
 at small $Q^2$  and  not very large $\nu$
(i.e., in  the  region where the  perturbative component can be neglected) are well described by
the nondiagonal GVDM [21-23].The sums over  vector meson  states in this  model  (analogous those in r.h.s. of Eq.(2.34)) contain  
no  cut-off functions, and, instead, the convergence of these sums is provided  by  large cancellations  between  the diagonal  and   nondiagonal 
transitions. This model, being essentially  nonperturbative, did not describe properly the region of very small $x_{Bj}$ 
values and therefore  must  be used  in  combination with models  describing the hard  component of structure functions.

The very simple and convenient for applications parameterization of numerical
calculations of the structure functions within the framework of the nondiagonal GVDM was suggested in the paper [5].  The corresponding
formulae are  given  in the Appendix  B.The simplicity of these parameterizations for the structure functions allows for
the appoximate analytical integration in the right-hand-side of Eq.(4.4).  
The result is:
$${v\frac{d\sigma}{dv}(E,v,m_l)\mid}_{nonpert}\cong\frac{\alpha}{2\pi}A\sigma_{\gamma p}(\nu)v^2
\{0.75G_{sh}(x)
[\varkappa ln(1+\frac{m^2_1}{Q^2_{min}})-\frac{\varkappa m^2_1}{m^2_1+Q^2_{min}}-$$
$$-\frac{2m^2_l}{Q^2_{min}}\frac{m_1^2+2Q_{min}^2}{m_1^2+Q_{min}^2}+\frac{4m_l^2}{m_1^2}ln(1+\frac{m_1^2}{Q_{min}^2})]
+0.25[(\varkappa+\frac{2m_l^2}{m_2^2}) ln(1+\frac{m^2_2}{Q^2_{min}})-$$
$$-\frac{2m^2_l}{Q^2_{min}}]+\frac{2\xi m^2_l}{Q^2_{min}}[0.75G_{sh}(x)\frac{m^2_1}{m^2_1+Q^2_{min}}
+0.25
\frac{m^2_2}{Q^2_{min}}ln(1+\frac{Q^2_{min}}{m^2_2})]\},     \eqno(4.6)$$
$$\varkappa\equiv 1-\frac2v+\frac2{v^2}. \eqno(4.6a)$$

This expression is written for a nuclear target. For the case of a nucleon target one must put $A=1,~~ G_{sh}=1$. The 
values of $m_1^2,m_2^2,$, and the definition of the function $G_{sh}(x)$ are given in the Appendix B.

The formula (4.6) differs from the corresponding result  of ref.[5] only  by several new  terms which  are  proportional
to ratios  $\frac{m_l^2}{m_1^2}$  and $\frac{m_l^2}{m_2^2}$. These new  terms  are negligibly  small  in  the  case
of the muon projectile, but become noticeable in the taon's case, especially in the limit of large $v$.   

The perturbative part of $v\frac{d\sigma}{dv}$ is obtained by substitution in the integral in Eq.(4.4)
 the structure functions given by Eqs.(2.15,2.16) with $(\stackrel{\,~-}{qq})p$ cross section $\sigma(r_\bot,s)$
given by Eq.(3.1) (we use the framework of the colour dipole model in the variant elaborated in [15]) or,
in unitarized form, by Eq.(3.6).

If the nonunitarized form of $\sigma(r_\bot,s)$ is used (supposing that the gluon saturation effects are small through 
the whole region of lepton energies considered in the present work (i.e., up to $10^9GeV$)), the energy dependence
 of $v\frac{d\sigma}{dv}$ is approximately factorized:
$${v\frac{d\sigma}{dv}(E,v,m_l)\mid}_{QCD}\simeq f(v,m_l)E^{\lambda_H}.    \eqno(4.7)$$
Here, $\lambda_H$ is the parameter entering the expressions (2.21,3.1). Correspondingly, in this case there is the power 
law rise with a lepton energy of the QCD part of the energy loss coefficient $b(E,m_l)$ defined by the integral
$$b(E,m_l)=N_A\int\limits^1_0 v\frac{d\sigma}{dv}(E,v,m_l)dv\sim E^{\lambda_H}.  \eqno(4.8)$$

The relative magnitude of the unitarization effects depends, in our phenomenological approach, on the several model parameters,
first of all on parameter $R$ which is supposed to be of the order of target size (see Eq.(3.5a)). The smaller parameter $R$, the sooner
logarithmic regime of the energy dependence sets in.

The sum of two contributions given in Eqs.(4.6) and (4.7) is, by terminology of cosmic ray physicists, the cross section 
of the photonuclear interaction:
$${v\frac{d\sigma}{dv}(E,v,m_l)=v\frac{d\sigma}{dv}(E,v,m_l)\mid}_{nonpert}+{v\frac{d\sigma}{dv}(E,v,m_l)\mid}_{QCD}.  \eqno(4.9)$$

The  values of parameters in the  expression (3.1)  for the  cross section $\sigma(r_\bot,s)$ must be determined from the
comparison  of  our model  predictions  with  experimental data for small $x_{Bj}$. The   example of such  a  comparison
for the  particular value $x_{Bj}=0.000032$ is  shown   on  fig.2. It  is  seen from  this figure that our GVDM  prediction  
for $F_2(x_{Bj},Q^2)$ is  clearly at  variance  with  data  even in   the region of relatively  low  values of $ Q^2$,  if  $x_{Bj}$
is small enough. At  the same time, one  can see   that   even  at very small $x_{Bj}$ there  is  a   $Q^2$-region,  in which
the hard component is  negligibly  small and  GVDM works well.  

The set  of  parameter  values  in  Eq.(3.1) found from this comparison and used   for  numerical  calculations  is  the  following:
  
$$~~~  a_1=0.99,~~~ a_2=0.7GeV, ~~~ a_3=-6.23GeV^2,$$ 

$$r_\bot^0=0.219GeV^{-1}, ~~~~ \lambda_H=0.387.        \eqno(4.10)$$

This set differs  from that obtaned  in   ref.[15] only  by the smaller value of the parameter $r_\bot^0$.

\section*{\bf \noindent \Large 5.Results of calculations}

We performed calculations of the photonuclear cross section and energy loss coefficient for muons and tau-leptons.

As is mentioned in the Introduction, muons and taus of super high energies ($\gtrsim10^5GeV$) are of natural origin: they appear as a 
result of interactions of superhigh energy cosmic rays in atmosphere (muons) and as a result of 
$\nu_{\mu}$( or $\nu_{e}$ )$\to\nu_{\tau}$ oscillations of extragalactic neutrinos on their way to the Earth with their subsequent interaction
somewhere near the (underground) detector (taus). 
Therefore, we need, in practice, in the case of superhigh energies of leptons, the cross section of the inelastic lepton scattering on a nuclear
target rather than on a nucleon one.

In the nonperturbative part of the cross section the nuclear shadowing effects are taken into account (formula (4.6)). As for the perturbative piece,
the problem of the nuclear shadowing requires a special investigation which will be performed in a separate work. Qualitatively, the
shadowing effects are small if 
the exponential term in the integral in the expression for the $(\stackrel{\,~-}{qq})Z$ interaction cross section,
$$\sigma_{\stackrel{\,~-}{qq}+Z}(r_\bot,s)=2\int d^2b\{1-exp[-\sigma(r_\bot,s)T(\stackrel{\to}{b})/2]\},   \eqno(5.1)$$
is close to 1. In this formula $T(\stackrel{\to}{b})$ is an optical thickness of the target nucleus at an impact parameter $\stackrel{\to}{b}$,
$$T(\stackrel{\to}{b})=\int \rho_A(b,z)dz.     \eqno(5.2)$$
The shadowing effects are small if
$$\sigma^2(r_\bot,s)\frac{T^2(\stackrel{\to}{b})}{8}<<\sigma(r_\bot,s)\frac{T(\stackrel{\to}{b})}{2},         \eqno(5.3)$$
or
$$\sigma_{max}(r_\bot,s)<<\frac{4}{T(\stackrel{\to}{b})}\lesssim\frac{4}{2R_A\rho_A}\sim\frac{8r^2_0}{A^{1/3}}.       \eqno(5.4)$$
Here, $\sigma_{max}$ is the maximum value of the $(\stackrel{\,~-}{qq})p$ interaction cross section,
$r_0\simeq1.2$fm. If $A=22$ (it is the value which we use in all numerical calculations in the present paper),
Eq.(5.4) gives the condition
$$\sigma_{max}(r_\bot,s)<<40mb.          \eqno(5.5)$$

The example of the calculation of $\sigma(r_\bot,s)$ is shown on fig.3, and it is seen that the condition (5.5) is violated only  
at very  high values  of $s$, $s>10^8GeV^2$. It  corresponds  to typical lepton energies $E=\nu  /v\sim s/(2M_p\cdot 0.1)\sim  10^9GeV $.   
Besides, the  condition (5.5) is unnecessarily strong and can be weakened by taking into account that the nuclear shadowing effect in perturbative part of 
$\sigma^A_{T,L}$ strongly depends on $Q^2$ because, due to properties of Bessel functions in the expressions for photon wave functions
(Eqs. 2.16), one has$$\stackrel{-}{Q}r_{\bot}\lesssim 1,          \eqno(5.6)$$
and, correspondingly,
$$r^2_{\bot char}\lesssim \frac{1}{Q^2},   \eqno(5.7) $$
where $r_{\bot char}$ is some characteristic value of a transverse size of the $\stackrel{\,~-}{qq}$-pair in the nonperturbative interaction.
The integral over $Q^2$ determining the photonuclear cross section $v\frac{d\sigma}{dv}$ is dominated by $Q^2$ values near $Q^2_{min}$
and, therefore, the role of  $\stackrel{\,~-}{qq}$-pairs with large $r_{\bot}$, for which the cross section $\sigma(r_{\bot},s)$ is relatively small,
is, due to Eq.(5.7), sufficient.Accordingly, shadowing  effects cannot be too large. 

Note  that the value of $r_\bot$ at  which  the cross section $\sigma(r_\bot,s)$   has a  maximum value does  not coincide (even 
approximately)  with a value of the  parameter  $r_\bot^0$ in the formula (3.1) (it is due  to  a large  value of the  parameter $a_3$,
leading  to  a fast reveal   of  the $a_3r^3$-term). So, the typical  pair's transverse size corresponding to the perturbative  interaction
is  determined, in  the   parameterization of  ref.[15], by the combination of  two parameters, $a_3$ and $r_\bot^0$.
 
The  results of our calculation of  the  perturbative part  of  $v\frac{d\sigma}{dv}$   are shown on  figs.4,5(for  $\mu  A$ and 
$\tau  A$ interactions).  These  results can
be used, together  with  the  corresponding nonperturbative parts  (given by Eq.(4.6)),  for  calculations of  a $\mu$ and  $\tau$
propagation  through thick  layers of matter  and for calculations of detection probabilities  of  superhigh  energy  leptons passing
through  large underground telescopes.  For completeness, we present on   figs.6,7  also the  nonperturbative  parts  of 
$v\frac{d\sigma}{dv}$. All calculations on   figs.4-7 are  done for standard rock, $A=22$.  
 
For a rough estimate of the photonuclear range of leptons in matter it  is  useful  to know the energy loss  coefficients,
$$-\frac{1}{E}\frac{dE}{dx}=b(E,m_l)=\frac{N_A}{A}\int v\frac{d\sigma}{dv}(E,v,m_l),           \eqno(5.8)$$
(in a case of the perturbative  part, if there is no shadowing, $v\frac{d\sigma}{dv}\sim A$ and we return to Eq.(4.8)). On figs.8,9
we show the results of calculations of such coefficients for standard rock. It is seen that at energies   
$E\sim10^8 - 10^9 GeV$ the contribution of  the  perturbative QCD part of  the total photonuclear  interaction in  the lepton
photonuclear energy losses is dominant ($\sim  70\%$).

The quantitatively similar results for the photonuclear energy loss coefficients $b(E,m_l)$ were obtained in ref.[1],
authors  of   which used for  the  description  of the structure functions ALLM[25] parameterization (based on the Regge
approach), and in ref.[26], for smaller energy interval ($E<10^6~GeV$) and only for muons (authors of [26] used for $\sigma_{T,L}$ 
the predictions of CKMT[27] model (also of Regge  type)). Needless to say, that the theoretical approaches of refs.[25,27]  are
completely different from the one presented here.

\section*{\bf \Large Acknowledgments}

Authors wish to thank  I. A. Sokalski for  the help  in numerical calculations.

\section*{\bf \Large  Appendix A: The cut-off functions}

In the case of transversely polarized virtual photons the criterion (2.14) and
Eqs.(2.3) lead to the  change (in the approximation $m_f=0$; $\Theta(x)$ is the step function )
$$\frac32\int\limits^{1}_{0}dz[z^2+(1-z)^2]\longrightarrow\frac32\int\limits^{1}_{0}dz[z^2+(1-z)^2\times$$
$$\times\Theta(\frac{k_{\bot 0}^2}
{M^2(1+\frac{Q^2}{M^2})^2}-z(1-z))
\Theta(\frac{k_{\bot 0}^2}{{M'}^2(1+\frac{Q^2}{{M'}^2})^2}-z(1-z))\equiv$$
$$\equiv\eta_T(M^2,{M'}^2,k_{\bot 0}^2).       \eqno(A1)$$
In diagonal approximation one obtains
$$\eta_T(M^2,M^2,k_{\bot 0}^2)\equiv\eta_T(M^2,k_{\bot 0}^2)=3z_0-3z^2_0+2z^3_0,           \eqno(A2)$$
$$z_0=\frac1 2-\sqrt{\frac14-\frac{k_{\bot 0}^2}{M^2(1+\frac{Q^2}{M^2})^2}}.     \eqno(A2a)$$
If $M^2>>k_{\bot 0}^2$ one has
$$z_0\cong\frac{k_{\bot 0}^2}{M^2(1+\frac{Q^2}{M^2})^2},~~\eta_T(M^2,k_{\bot 0}^2)\cong 3\frac{k_{\bot 0}^2}
{M^2(1+\frac{Q^2}{M^2})^2}.     \eqno(A3)$$

Having in mind that nonperturbative $(\stackrel{\,~-}{qq})p$ interaction amplitudes are essentially diagonal  (it is known  that, in 
the  black disc  limit, at least, the   cross sections of nondiagonal diffractive processes should be small), we can 
approximate $\eta_T(M^2,{M'}^2,k_{\bot 0}^2)$ by the formula 
$$\eta_T(M^2,{M'}^2,k_{\bot 0}^2)\cong\sqrt{\eta_T(M^2,k_{\bot 0}^2)\eta_T({M'}^2,k_{\bot 0}^2)}.        \eqno(A4)$$

In the case of longitudinally polarized photons the analogue of Eq.(A1) is 
$$\frac32\int\limits^{1}_{0}dz4z(1-z)\longrightarrow\frac32\int\limits^{1}_{0}dz4z(1-z)\Theta(\frac{k_{\bot 0}^2}{M^2(1+
\frac{Q^2}{M^2})^2}-z(1-z))\times$$
$$\times\Theta(\frac{k_{\bot 0}^2}{{M'}^2(1+\frac{Q^2}{{M'}^2})^2}-z(1-z))\equiv\eta_L(M^2,{M'}^2,k_{\bot 0}^2),    \eqno(A5)$$
and, in diagonal approximation
,
$$\eta_L(M^2,M^2,k_{\bot 0}^2)\equiv\eta_L(M^2,k_{\bot 0}^2)=6z^2_0-4z^3_0.               \eqno(A6)$$
Similarly with the previous case we approximate $\eta_L(M^2,{M'}^2,k_{\bot 0}^2)$ by the formula
$$\eta_L(M^2,{M'}^2,k_{\bot 0}^2)\cong\sqrt{\eta_L(M^2,k_{\bot 0}^2)\eta_L({M'}^2,k_{\bot 0}^2)}.                     \eqno(A7)$$

\section{\bf \Large Appendix B: Parameterization of structure functions in GVDM}

In  the GVDM version of  ref.[5],  the electromagnetic structure functions of a proton are well parameterized  
by  the  following  expressions:
$$\sigma_{T}(Q^2,\nu)=\sigma_{\gamma p}(\nu)\left\{0.75\frac{m^4_1}{(m^2_1+Q^2)^2}+0.25\frac{m^2_2}{m^2_2+Q^2}\right\}, \eqno(B1)$$
$$\sigma_L(Q^2.\nu)=\xi\sigma_{\gamma p}(\nu)\{0.75\frac{m^2_1Q^2}{(m^2_1+Q^2)^2}+$$
$$+0.25[\frac{m^2_2}{Q^2}ln(1+\frac{Q^2}{m^2_2})-\frac{m^2_2}{m^2_2+Q^2}]\},   \eqno(B2)$$
$$m^2_1=0.54~ GeV^2,~~m^2_2=1.8~GeV^2,~~\xi=0.25.        \eqno(B3)$$
In these formulae $\sigma_{\gamma p}(\nu)$ is the total photoproduction cross section for a real photon of energy $\nu$, which was parametrized
by the equation   [5]
$$\sigma_{\gamma p}(\nu)=114.3+1.647ln^2(0.0213\nu)~\mu b     \eqno(B4)$$
($\nu$ is in units of $GeV$). The parameterization (B1,B2) works rather well in the small $Q^2-$region 
($Q^2\lesssim1~GeV^2$) and  in the region of not very small $x_{Bj}$ ($10^{-3}<x_{Bj}<10^{-1}$), and does 
not contradict with the most recent H1 and ZEUS data.

The generalization of Eqs.(B1,B2) for nuclear targets is the following [5]: one must introduce the common factor $A$ and multiply the first
term in figure brackets in right-hand-sides in Eqs.(B1,B2) by the shadowing function $G_{sh}(x)$ defined by the relation 
$$G_{sh}(x)=\frac{3}{x^3}(\frac{x^2}{2}-1+e^{-x}(1+x)),         \eqno(B5)$$
$$x\equiv0.00282A^{1/3}\sigma_{\gamma p}(\nu).        \eqno(B5a)$$
As  a result,  one  obtains

$$\sigma^{A}_T(Q^2,\nu)=A\sigma_{\gamma p}(\nu)\{0.75\frac{m^4_1}{(m^2_1+Q^2)^2}G_{sh}(x)+
0.25\frac{m^2_2}{m^2_2+Q^2}\}, \eqno(B6)$$
$$\sigma^{A}_L(Q^2,\nu)=A\sigma_{\gamma p}(\nu)\xi\{0.75\frac{m^2_1Q^2}{(m^2_1+Q^2)^2}G_{sh}(x)+$$
$$+0.25[\frac{m^2_2}{Q^2}ln(1+\frac{Q^2}{m^2_2})-
\frac{m^2_2}{m^2_2+Q^2}]\}.      \eqno(B7) $$

\newpage
\begin{figure}
\centerline{\includegraphics{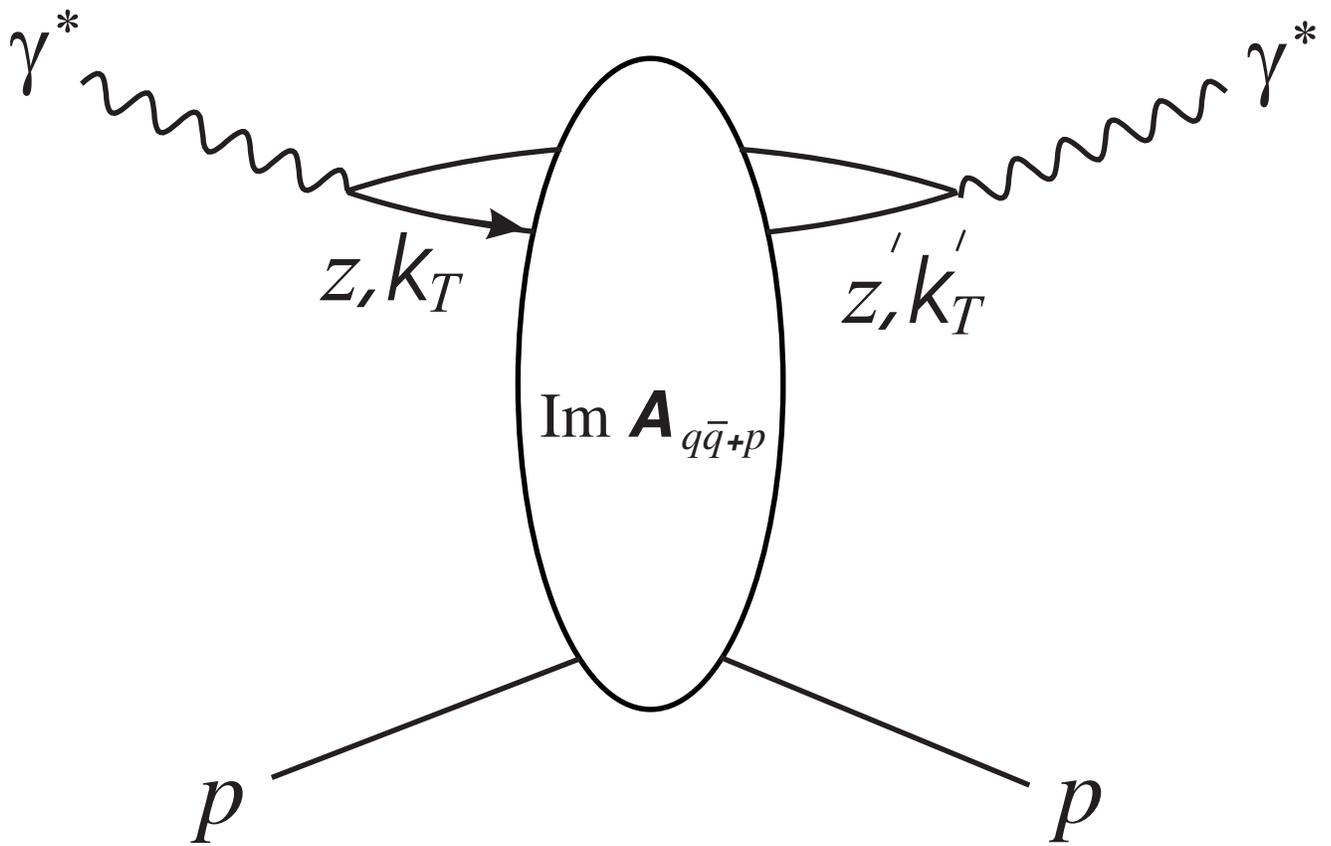}}
\caption { The  diagram representing schematically the contribution  of  $\protect\stackrel{\,~-}{qq}$-channel  in  the imaginary
part of the Compton forward scattering amplitude. }\label{Fi:1}
\end{figure}

\newpage
\begin{figure}
\centerline{\includegraphics{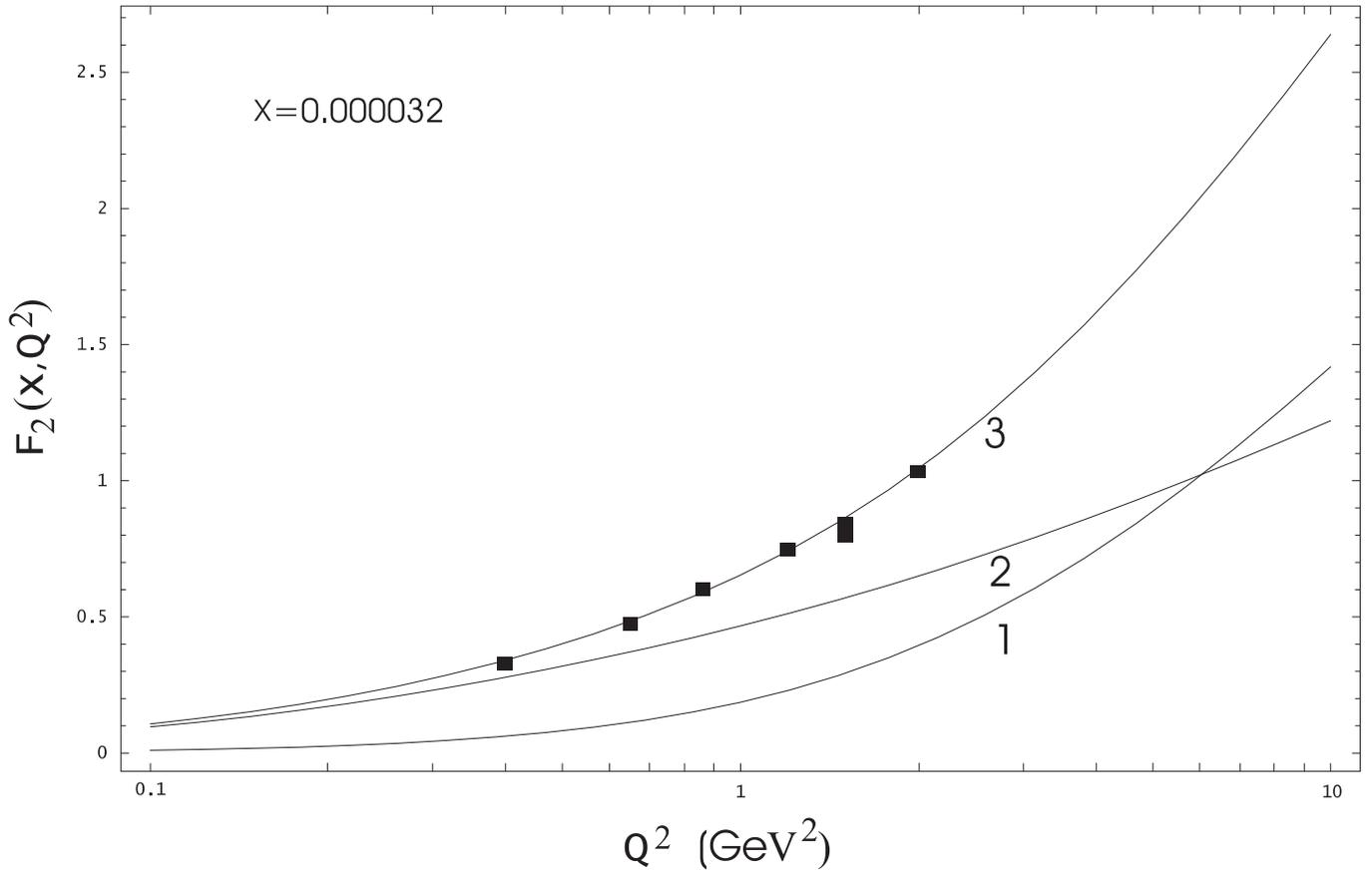}}
\caption { The proton structure function  $F_2$,  for  the fixed  value   of $x_{Bj}$,  as a function of  $Q^2$. 1: hard component
 (the set of parameters, given   by Eq.(4.10), is  used); 2:  GVDM part (formulae  (B1) and (B2) are  used); 3:  summary curve.
 Experimental points are taken from  ref.[24]. }\label{Fi:2}
\end{figure}

\newpage
\begin{figure}
\centerline{\includegraphics{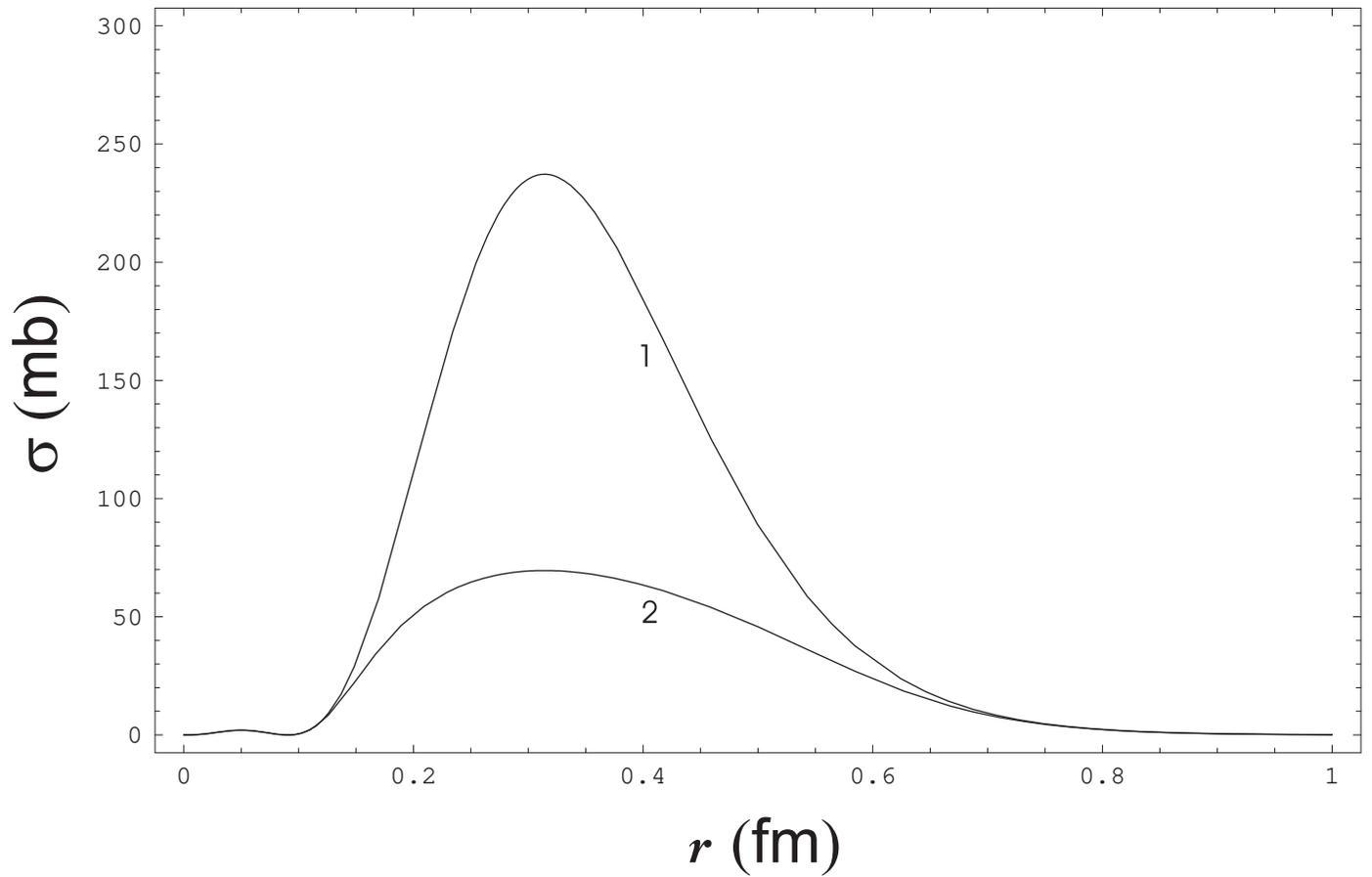}}
\caption { The  total  cross section  of an interaction of the $\stackrel{\,~-}{qq}$-pair of the transverse  size  $r_\bot$ with
a  target proton, $\sigma(r_\bot,s)$. 1:  Eq.(3.1);  2: the unitarized  cross section (Eq.(3.6)). $s=10^8GeV^2$ in  both   cases. }\label{Fi:2}
\end{figure}

\newpage
\begin{figure}
\centerline{\includegraphics{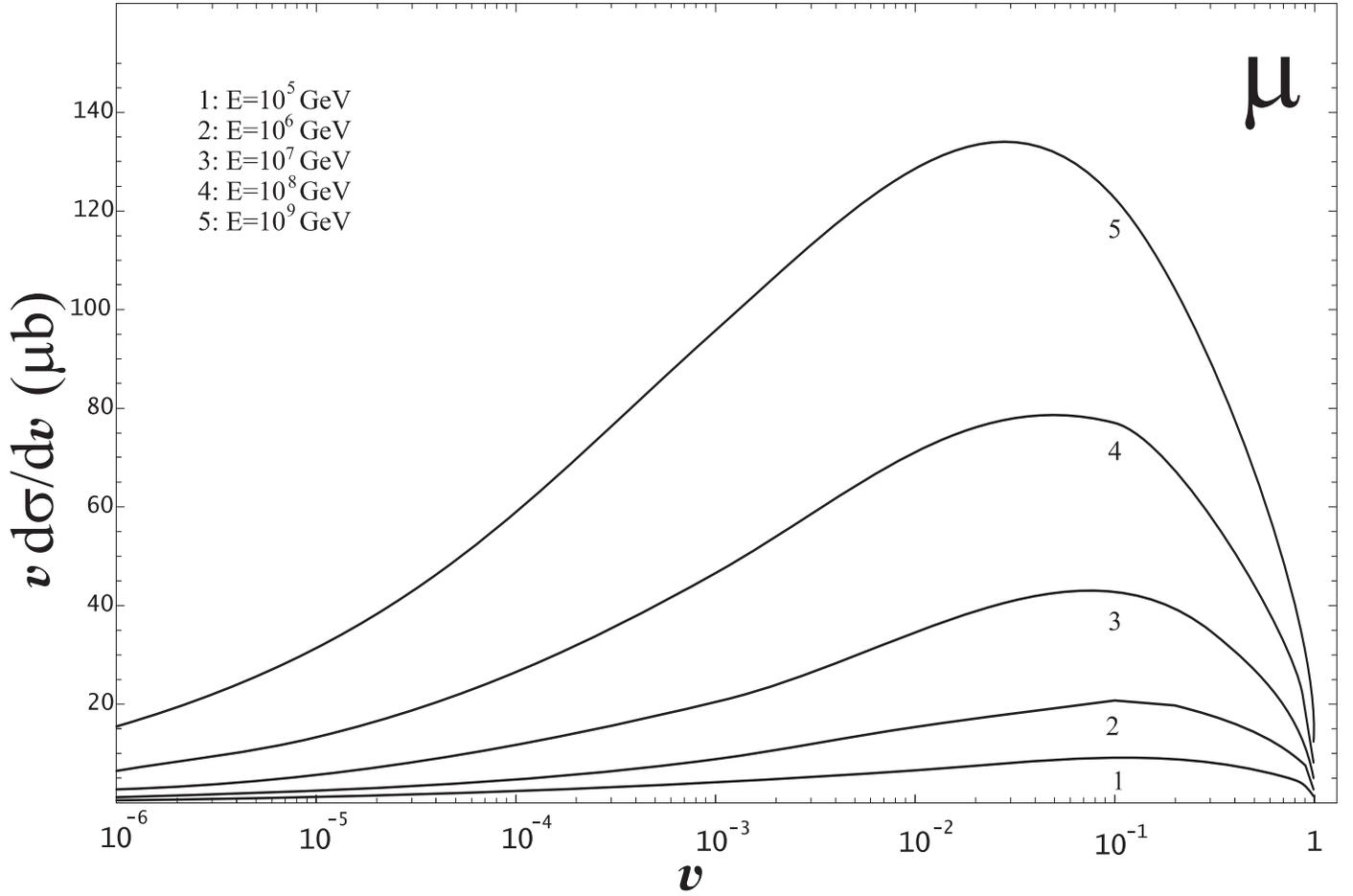}}
\caption{ The perturbative part of the differential photonuclear ($\mu A$) cross section for $A=22$ (integrated over $Q^2$),
as a  function of   $v$, the fraction  of  the muon energy   transferred to  hadrons, for  several values of 
an initial energy  of the   muon. The expressions (4.4), (2.15), (2.16), (3.6), and the set of parameters (4.10) are used 
for the calculation.}\label{Fi:3}
\end{figure}

\newpage
\begin{figure}
\centerline{\includegraphics{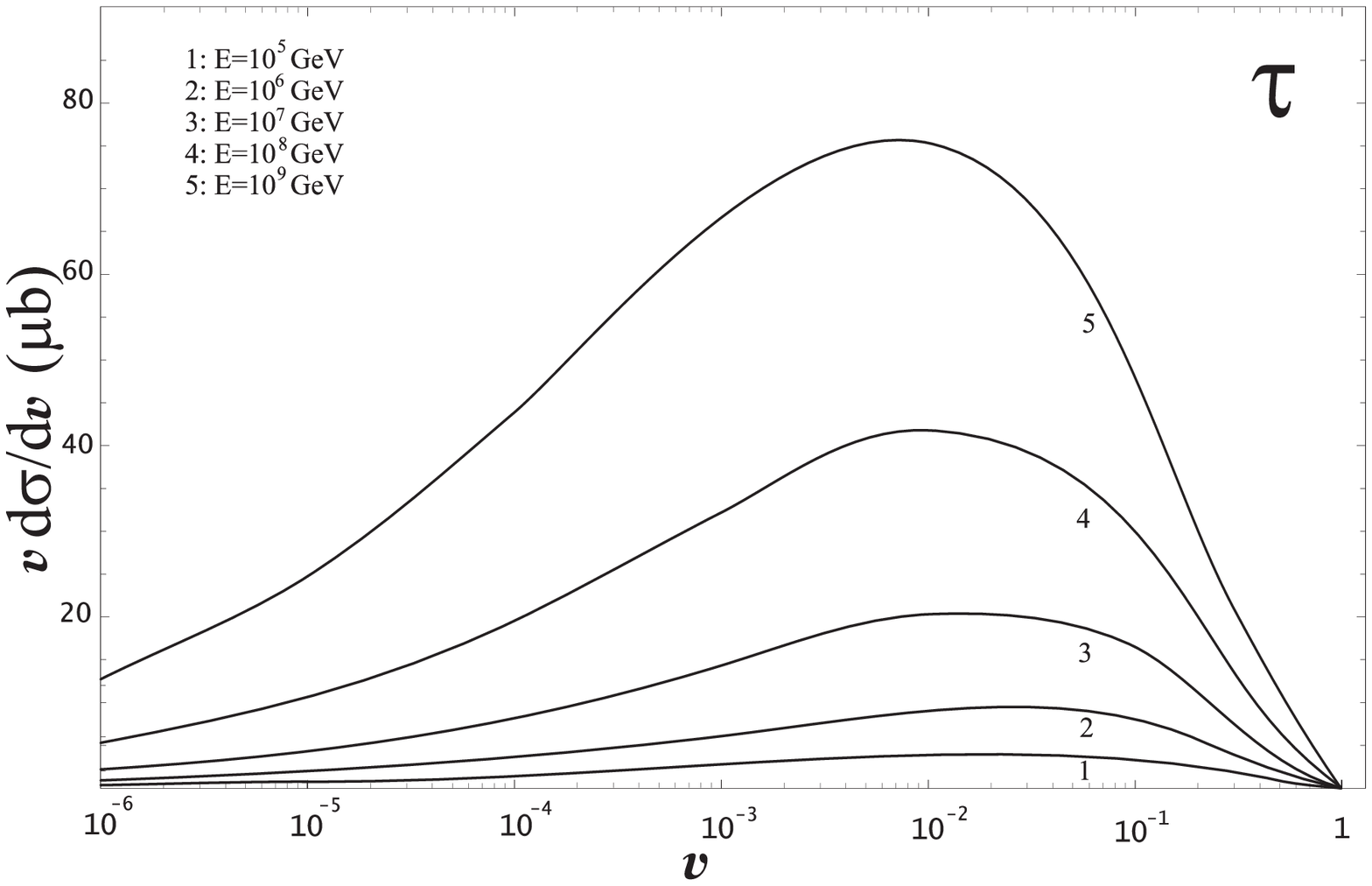}}
\caption { The  same  as fig.4, except  the   change $\mu \to \tau $.}\label{Fi:4}
\end{figure}

\newpage
\begin{figure}
\centerline{\includegraphics{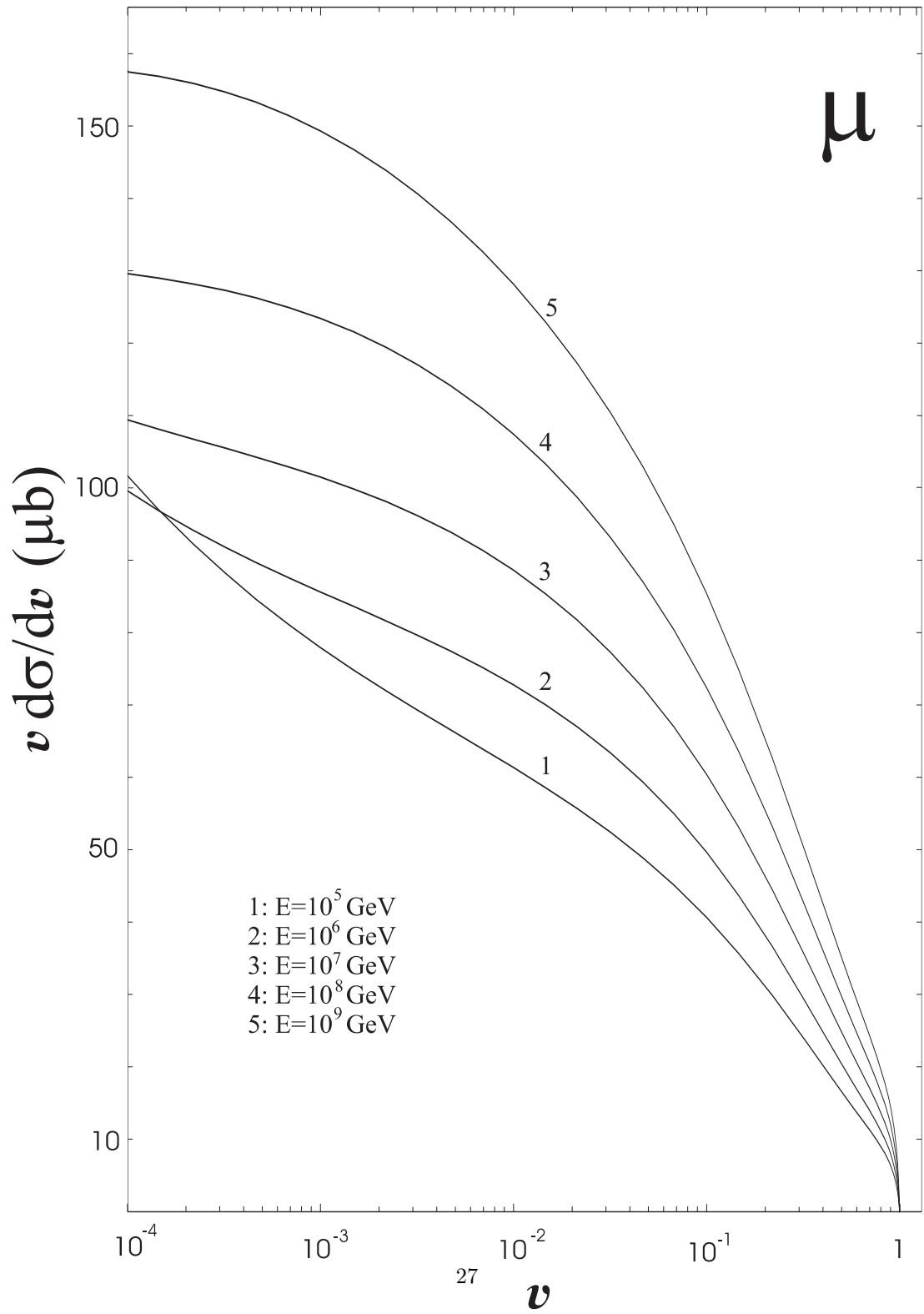}}
\caption {The nonperturbative part of the differential photonuclear ($\mu A$) cross section calculated using Eq.(4.6), 
for $A=22$.}\label{Fi:5}
\end{figure}

\newpage
\begin{figure}
\centerline{\includegraphics{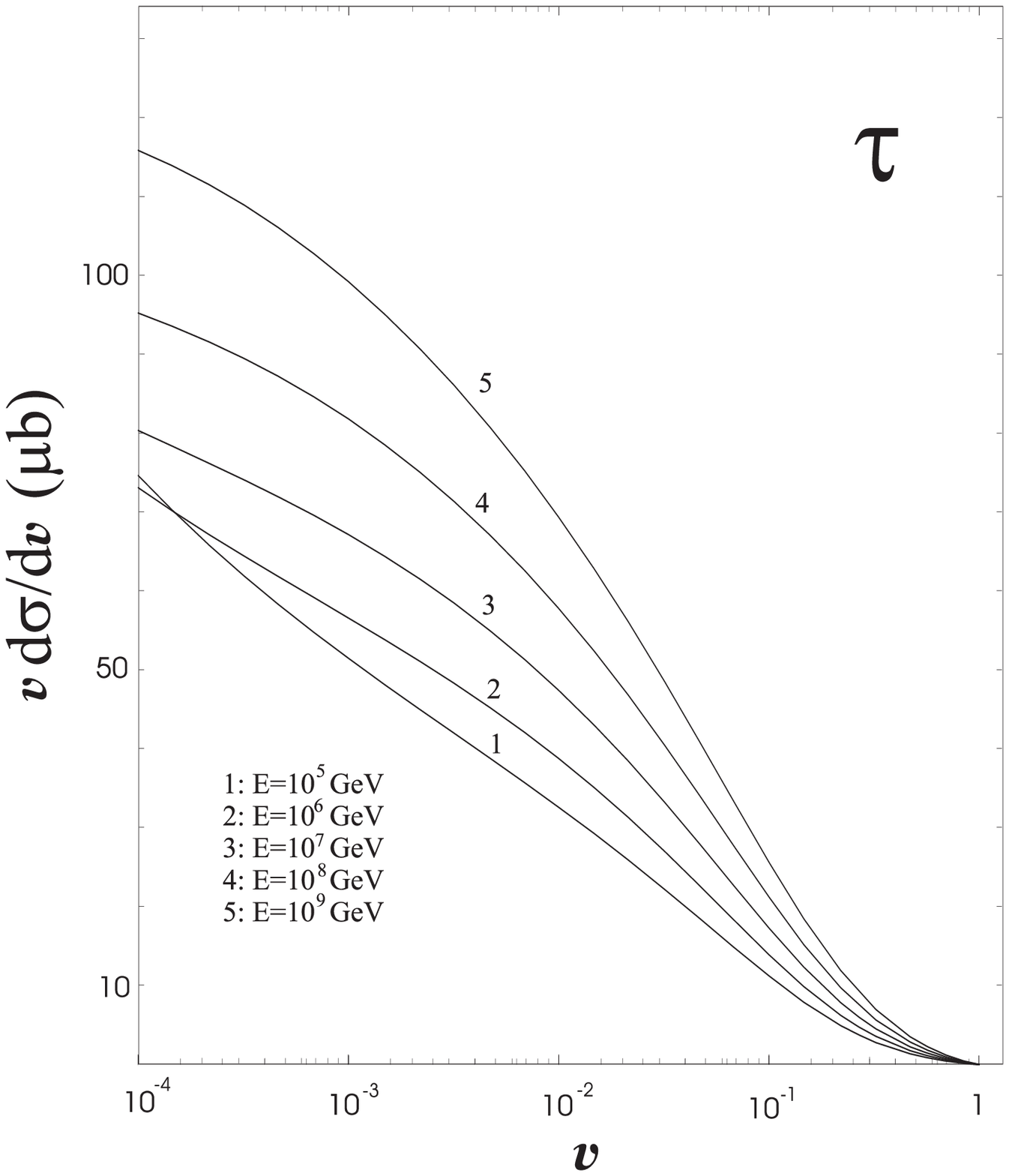}}
\caption { The same   as fig.6, except the change $\mu \to \tau$. }\label{Fi:6}
\end{figure}

\newpage
\begin{figure}
\centerline{\includegraphics{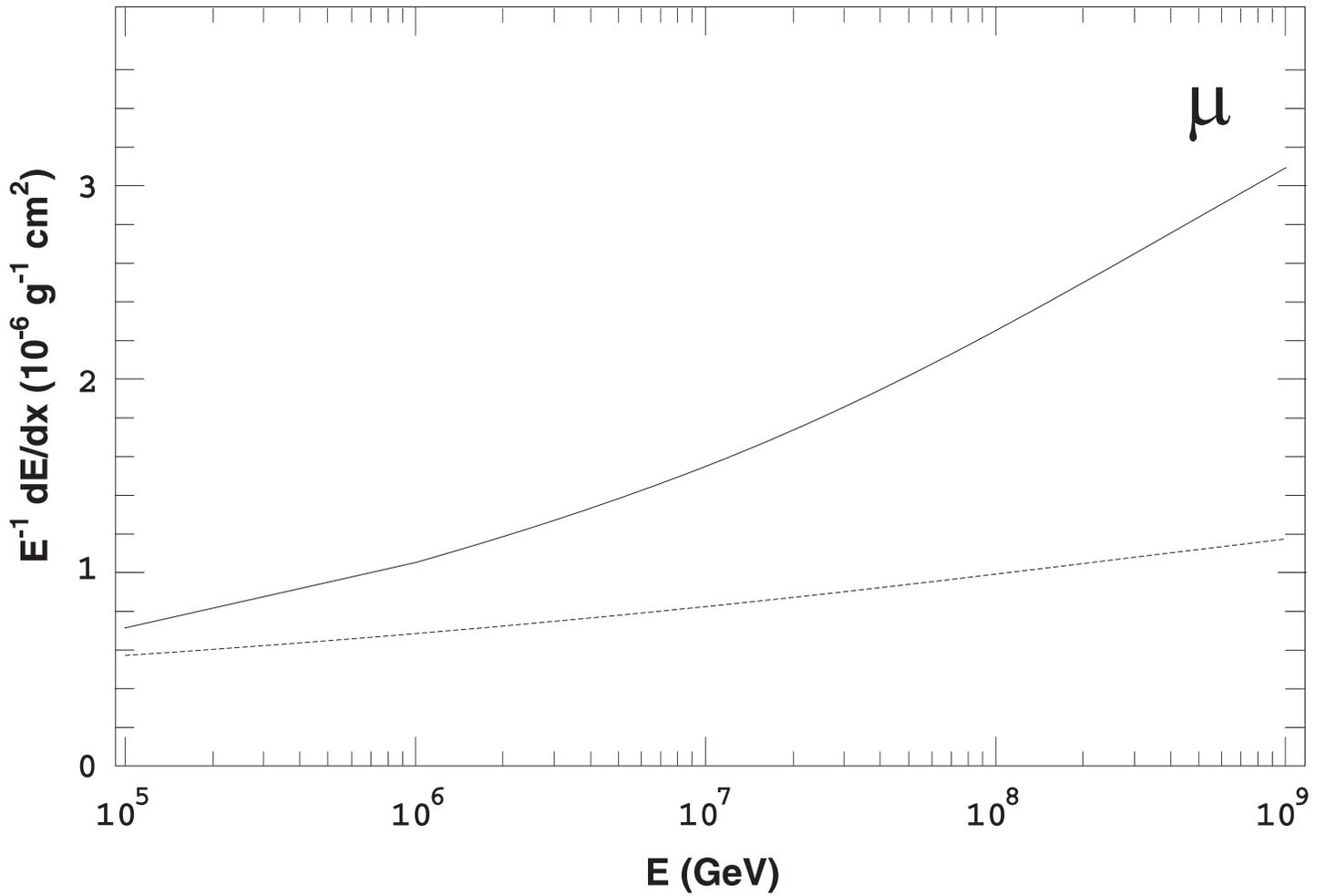}}
\caption{ The energy  loss coefficient  for a muon, as a function  of  its energy, for standard rock($A=22$).  The solid
line is the total coefficient, the dashed line is the nonperturbative part calculated using BB formulae [5].}\label{Fi:7}
\end{figure}

\newpage
\begin{figure}
\centerline{\includegraphics{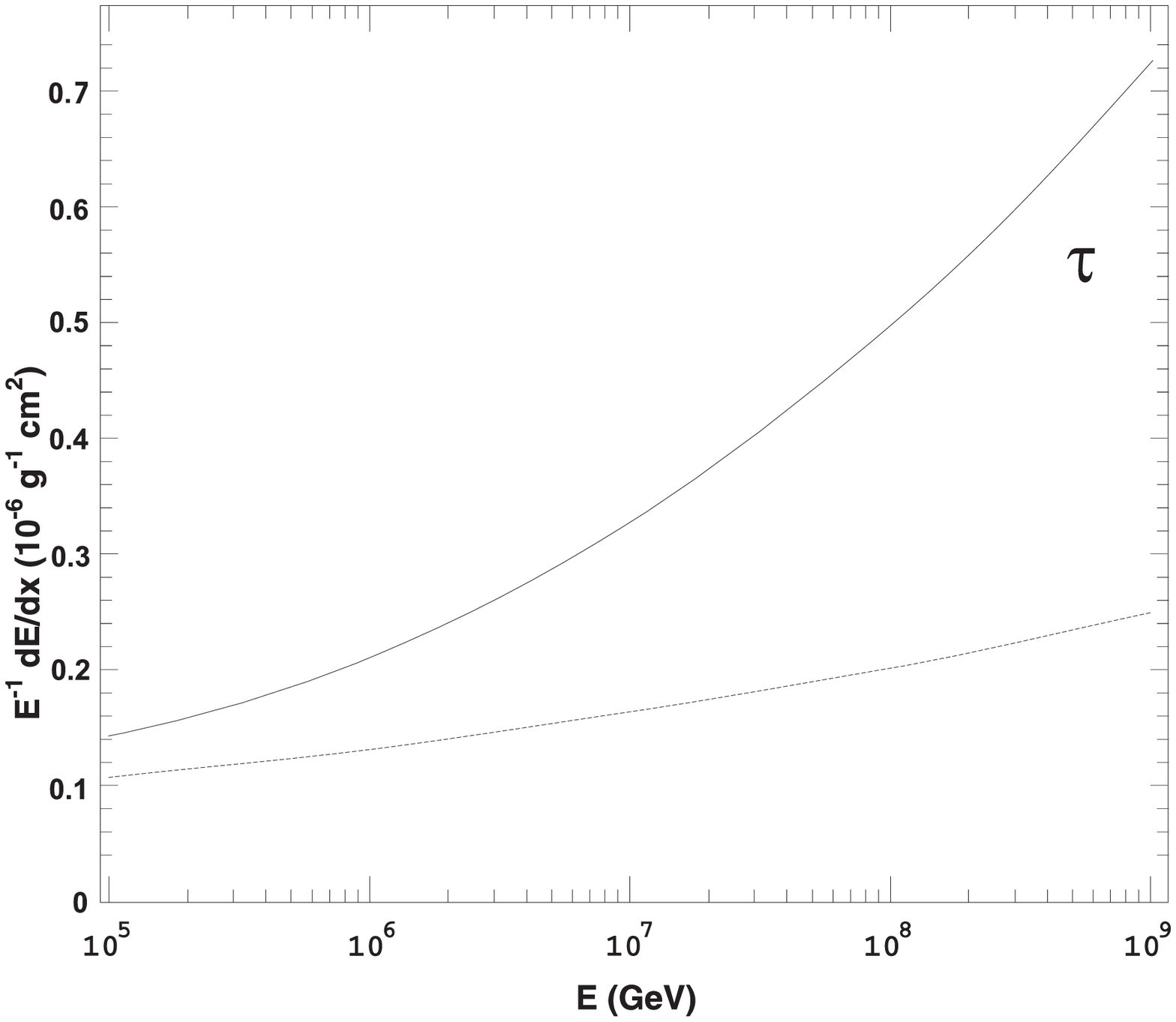}}
\caption {The same as fig.8, except  the change $\mu \to \tau $.}\label{Fi:8}
\end{figure}

\end{document}